\documentclass[conference]{IEEEtran}
\usepackage{color}
\usepackage{array}
\usepackage{cite}
\usepackage{multirow}
\usepackage{stmaryrd}
\usepackage{setspace}
\usepackage{amssymb}
\usepackage[cmex10]{amsmath}
\usepackage{amsthm}
\usepackage[pdftex]{graphicx}
\usepackage{epstopdf}
\usepackage{epsfig}
\usepackage{verbatim}
\usepackage{bbding}
\usepackage{dsfont}
\newcommand{\ma}[1]{\ensuremath\mathcal{#1}}
\newcommand{\bs}[1]{\ensuremath\boldsymbol{#1}}
\newcommand{\ds}{\ensuremath\displaystyle\sum}

\newsavebox{\subfigure}
\newsavebox{\theorembox}
\newsavebox{\lemmabox}
\newsavebox{\corollarybox}
\newsavebox{\propositionbox}
\newsavebox{\examplebox}
\newsavebox{\conjecturebox}
\newsavebox{\algbox}
\newsavebox{\qbox}
\newsavebox{\problembox}
\newsavebox{\definitionbox}
\newsavebox{\assumptionbox}
\newsavebox{\hypothesisbox}
\newsavebox{\factbox}
\newsavebox{\remarkbox}
\savebox{\theorembox}{\noindent\bf Theorem}
\savebox{\lemmabox}{\noindent\bf Lemma}
\savebox{\corollarybox}{\noindent\bf Corollary}
\savebox{\propositionbox}{\noindent\bf Proposition}
\savebox{\examplebox}{\noindent\bf Example}
\savebox{\conjecturebox}{\noindent\bf Conjecture}
\savebox{\algbox}{\noindent\bf Algorithm}
\savebox{\qbox}{\noindent\bf Question}
\savebox{\definitionbox}{\noindent\bf Definition}
\savebox{\problembox}{\noindent\bf Problem}
\savebox{\assumptionbox}{\noindent\bf Assumption}
\savebox{\hypothesisbox}{\noindent\bf Hypothesis}
\savebox{\factbox}{\noindent\bf Fact}
\savebox{\remarkbox}{\noindent\bf Remark}
\newtheorem{theorem}{\usebox{\theorembox}}

\begin{document}
\title{\huge Density Evolution Analysis of Node-Based Verification-Based Algorithms in Compressed Sensing}
\author{Yaser Eftekhari, Anoosheh Heidarzadeh, Amir H. Banihashemi, Ioannis Lambadaris\\
			Carleton University, Department of Systems and Computer Engineering, Ottawa, ON, Canada\\
			E-mails: \{eft-yas, anoosheh, Amir.Banihashemi, Ioannis\}@sce.carleton.ca}
\maketitle
\thispagestyle{empty}
%\newpage
%\linenumbers
%\linenumbersep 30pt\relax
%\linenumbersep 30pt
\begin{abstract}
In this paper, we present a new approach for the analysis of iterative node-based verification-based (NB-VB) recovery algorithms in the context of compressive sensing. These algorithms are particularly interesting due to their low complexity (linear in the signal dimension $n$). The asymptotic analysis predicts the fraction of unverified signal elements at each iteration $\ell$ in the asymptotic regime where $n \rightarrow \infty$. The analysis is similar in nature to the well-known density evolution technique commonly used to analyze iterative decoding algorithms. To perform the analysis, a message-passing interpretation of NB-VB algorithms is provided. This interpretation lacks the extrinsic nature of standard message-passing algorithms to which density evolution is usually applied. This requires a number of non-trivial modifications in the analysis. The analysis tracks the average performance of the recovery algorithms over the ensembles of input signals and sensing matrices as a function of $\ell$. Concentration results are devised to demonstrate that the performance of the recovery algorithms applied to any choice of the input signal over any realization of the sensing matrix follows the deterministic results of the analysis closely. Simulation results are also provided which demonstrate that the proposed asymptotic analysis matches the performance of recovery algorithms for large but finite values of $n$. Compared to the existing technique for the analysis of NB-VB algorithms, which is based on numerically solving a large system of coupled differential equations, the proposed method is much simpler and more accurate.
\end{abstract}
%%%%%%%%%%%%%%%%%%%%%%%%%%%%%%%%%%%%%%%%%%%%%%%%%%%%%%%%%%%%%%%%%%%%%%%%%%%%%%%%%%%%%%%%%%%%%%%%%%%%%%%%%%%%%
%%%%%%%%%%%%%%%%%%%%%%%%%%%%%%%%%%%%%%%%%%%%%%%%%%%%%%%%%%%%%%%%%%%%%%%%%%%%%%%%%%%%%%%%%%%%%%%%%%%%%%%%%%%%%
\section{Introduction}
Noiseless compressed sensing was introduced with the idea to represent a signal $\bs{v}\in\mathbb{R}^n$ with $k$ non-zero elements using measurements $\bs{c}\in\mathbb{R}^m$, where $k<m\ll n$, and yet be able to recover the original signal $\bs{v}$ back \cite{D06,CRTFeb06}. In the measuring process, also referred to as \textit{encoding}, signal elements are mapped to measurements through a linear transformation represented by the matrix multiplication $\bs{c} = \bs{G}\bs{v}$, where the matrix $\bs{G}\in \mathbb{R}^{m\times n}$ is referred to as the \emph{sensing matrix}. This linear mapping can also be characterized by a bipartite graph \cite{XH07}, referred to as the \textit{sensing graph}.

In the recovery process, also referred to as \textit{decoding}, based on the knowledge of the measurements and the sensing matrix, we estimate the original signal. The decoding process is successful if $\bs{v}$ is estimated correctly. Three performance measures namely, \emph{density ratio} $\gamma \triangleq k/n$, \emph{compression ratio} $r_c \triangleq n/m$, and \emph{oversampling ratio} $r_o \triangleq m/k$ are used in order to measure and compare the performance of the recovery algorithms in the context of compressed sensing.\footnote{In \cite{WV09} authors proved that the lower bound $r_o = 1$ ($\gamma = 1/r_c$) is achievable in the asymptotic case ($n \rightarrow \infty$). Indeed, a decoder based on the $\ell_0$ norm can achieve $r_0 \approx 1$ under certain conditions \cite{B07}. Despite its conspicuous performance, the solution to the $\ell_0$ recovery is numerically unstable \cite{T04}.}

Perhaps one of the most interesting recovery algorithms in compressed sensing, and also a suitable choice for low-complexity encoding/decoding purposes, is a sub-class of iterative message-passing algorithms called \textit{Verification-Based} (VB) algorithms. The two algorithms in this class were introduced in the context of channel coding over non-binary alphabet \cite{LM05}, adopted in the context of compressed sensing \cite{SBB206,ZP09}, and studied in numerous literature \cite{LM05, ZP07J, ZP08, LMPDK08, APT10, SBB206, ZP09, XH07}. For each VB algorithm two descriptions exist, which are known as \emph{node-based (NB)} and \emph{message-based (MB)} \cite{ZP07J,ZP09}.\footnote{In the context of turbo codes, NB and MB approaches are known as non-extrinsic and extrinsic message-passing, respectively \cite{BG96}.} The VB algorithms are well-known for their low computational complexity, $O(n)$, and their insensitivity to the distribution of non-zero entries in both the sensing matrix and the input signal (under certain conditions). These algorithms are, however, expected to be sensitive to the presence of noise in the measurements. Therefore, the performance of noise-free algorithms can serve as an upper bound for the performance of the noisy versions in real-life applications.

Authors in \cite{LM05,ZP07J,ZP09} demonstrated that for each VB algorithm as $n \rightarrow \infty$, there exists a limiting value for $\gamma$, before which the recovery algorithm is successful with probability one. Researchers in \cite{LM05,ZP09,ZP07J} analyzed VB algorithms in the asymptotic case to calculate this limiting value, henceforth referred to as \textit{success threshold}, and then use it to compare different VB algorithms and estimate their performance for finite (large) $n$. The NB and MB descriptions yield different success thresholds\footnote{For one of the VB algorithms, the two descriptions perform the same, while for the other one, the NB version outperforms the MB one \cite{ZP07J}.} and are analyzed using different tools. In general, NB descriptions have higher success thresholds and are harder to analyze. Asymptotic analysis considered in \cite{LM05,ZP09} are of MB type, while authors in \cite{ZP07J} considered the NB type for the recovery algorithms. In this paper, we generalize/revise the density evolution analysis \cite{RU01} and propose a low-complexity analysis for the NB descriptions.
%%%%%%%%%%%%%%%%%%%%%%%%%%%%%%%%%%%%%%%%%%%%%%%%%%%%%%%%%%%%%%%%%%%%%%%%%%%%%%%%%%%%%%%%%%%%%%%%%%%%%%%%%%%%%
%%%%%%%%%%%%%%%%%%%%%%%%%%%%%%%%%%%%%%%%%%%%%%%%%%%%%%%%%%%%%%%%%%%%%%%%%%%%%%%%%%%%%%%%%%%%%%%%%%%%%%%%%%%%%
\section{Previous Works and Motivation}
Density evolution \cite{RU01} is a well-known method to analyze iterative message-passing algorithms in coding theory. In this method, the probability of some events of interest are calculated and tracked with iteration number. In the case of MB decoders, the calculation and evolution of such probabilities follow a closed form solution discussed in \cite{LM05}. Due to the fundamental difference between the NB and MB decoders, the existing density evolution methodology can not be used to analyze the performance of NB algorithms \cite{ZP07J}. For this reason, authors in \cite{ZP07J} used techniques based on differential equations for formulating the density evolution for the NB versions. Due to the large number of resulting coupled differential equations, the authors used numerical methods to evaluate the performance of the NB algorithms. The problems associated with their approach are: 1) since the approximation depends on the choice of $n$, and the analysis is valid for $n \rightarrow \infty$, one has to choose very large $n$, which directly translates to long running time and high computational complexity, and 2) the errors in the numerical approximations, potentially, propagate through iterations. Hence, there is no guarantee on the reported success threshold by this analysis, even for large $n$, as our numerical results show in Section \ref{simulation}.

In this paper we first define the ensemble of graphs and inputs of interest as well as the \emph{state parameters} that fully describe the state of the NB-VB decoders at any iteration. Then, similar to \cite{RU01}, we show that for a given decoder, at any iteration, the state parameters associated with almost all realizations of graphs and inputs are concentrated around the state parameters calculated using a deterministic algorithm (also referred to as the {\em density evolution analysis}). Using the density evolution analysis, we can estimate the success threshold of different VB algorithms over a variety of sparse graphs. The proposed analysis in this paper assumes noiseless measurements and has a computational complexity linear in the number of iterations. Compared to the differential equation analysis in \cite{ZP07J}, our approach uses almost the same number of state parameters but relies on addition and multiplication operations resulting in substantially less complexity.
%%%%%%%%%%%%%%%%%%%%%%%%%%%%%%%%%%%%%%%%%%%%%%%%%%%%%%%%%%%%%%%%%%%%%%%%%%%%%%%%%%%%%%%%%%%%%%%%%%%%%%%%%%%%%
%%%%%%%%%%%%%%%%%%%%%%%%%%%%%%%%%%%%%%%%%%%%%%%%%%%%%%%%%%%%%%%%%%%%%%%%%%%%%%%%%%%%%%%%%%%%%%%%%%%%%%%%%%%%%
\section{Background}
\label{back_knowledge}
%%%%%%%%%%%%%%%%%%%%%%%%%%%%%%%%%%%%%%%%%%%%%%%%%%%%%%%%%%%%%%%%%%%%%%%%%%%%%%%%%%%%%%%%%%%%%%%%%%%%%%%%%%%%%
\subsection{Ensembles of Sensing Graphs and Inputs}
\label{Defs}
A $(d_v,d_c)$-\textit{regular bipartite weighted graph} (or \textit{weighted biregular graph}) $\ma{G}({V}\cup{C},W(E))$ is a graph whose set of vertices ${V}\cup{C}$ can be divided into two disjoint sets ${V}$ and ${C}$, so that every (weighted) edge in the set ${E}$ connects a vertex in ${V}$ to one in ${C}$. Moreover, each vertex in $V$ ($C$) is incident to $d_v$ ($d_c$) vertices in $C$ ($V$), where $d_v$ and $d_c$ are fixed positive integers. Following the terminology of graph codes, we refer to the sets $V$ and $C$ as the \textit{variable nodes} and \textit{check nodes}, respectively. To each such graph, corresponds a biadjacency matrix $\bs{A}(\ma{G})$ of size $|{C}|\times |{V}|$ formed as follows: assuming an arbitrary labeling for nodes in $V$ and $C$, if there exists an edge between the vertices $c_i\in C$ and $v_j\in V$, then the entry $a_{ij}$ in the matrix is the weight of such edge, otherwise, $a_{ij} = 0$. Any such biadjacency matrix has $n - d_c$ $0$'s in each row and $m - d_v$ $0$'s in each column.

For given parameters $d_v,d_c$ and $n$ ($m=n d_v/d_c$), let $\ma{W}_{f}^{n}(d_v,d_c)$ denote the ensemble of biadjacency matrices of interest corresponding to all $(d_v,d_c)$-weighted biregular graphs with $n$ variable nodes, whose weights $w$ are drawn i.i.d. according to a distribution $f(w)$.

To discuss the ensemble of inputs of interest, let $\alpha\in [0,1]$ be a fixed real number and $\bs{v}$ be a vector of length $n$ with elements $\bs{v}_i$ drawn i.i.d. according to a mixed probability distribution function defined as follows: the element is zero with probability $1-\alpha$, or follows a distribution $g$ with probability $\alpha$. We denote the ensemble of all such vectors by $\ma{V}_{g}^{n}(\alpha)$.

In this paper we consider weighted biregular sensing graphs.\footnote{It is part of our ongoing research to extend the results of this paper to (weighted) irregular graphs. Our preliminary simulations show that the success threshold associated with VB algorithms can be increased by the use of carefully designed irregular graphs.} We map the sets of signal elements and measurements to the vertex sets $V$ and $C$ ($|V|=n$, $|C|=m$), respectively. We will interchangeably use the terms variable nodes and signal elements as well as check nodes and measurements. Also, the sensing matrix $\bs{G}$ is regarded as a biadjacency matrix drawn at random from the ensemble $\ma{W}_f^{n}(d_v,d_c)$. Moreover, we draw the signal vector $\bs{v}$ at random from the ensemble $\ma{V}_{g}^{n}(\alpha)$.
%%%%%%%%%%%%%%%%%%%%%%%%%%%%%%%%%%%%%%%%%%%%%%%%%%%%%%%%%%%%%%%%%%%%%%%%%%%%%%%%%%%%%%%%%%%%%%%%%%%%%%%%%%%%%
\subsection{VB Algorithms and Verification Rules}
\label{VBenc}
The first VB algorithm discussed in this paper, referred to as LM, is the LM1 algorithm in \cite{ZP09}. The second algorithm, referred to as SBB, is the ``sudocode'' algorithm in \cite{SBB206}, which is the same as the LM2 algorithm in \cite{ZP09}. In these algorithms, a variable node can be either: ``verified'' or ``unverified''. Based on the following three verification rules, variable nodes are verified and certain values are assigned to them. 
These values remain unchanged throughout subsequent iterations.
\begin{itemize}
	\item Zero Check Node (ZCN): If a check node has a value equal to zero, all its neighboring variable nodes are verified with a value equal to zero.
	\item Degree One Check Node (DOCN): If a check node is connected to only one unverified variable node, the variable node is verified with the value of the check node.
	\item Equal Check Nodes (ECN): If $N$ check nodes have the same non-zero value then 1) all variable nodes neighbor to a subset (not all) of such check nodes are verified with the value zero, and 2) if there exists a unique variable node neighbor to all $N$ check nodes, then it is verified with the common value of the check nodes.
\end{itemize}

The SBB algorithm applies all verification rules, while the LM algorithm applies all but the ECN rule. In this (non message-passing) description, when a variable node is verified at an iteration, its verified value is subtracted from the value of its neighboring check nodes, and then, removed from the sensing bigraph along with all its adjacent edges. Verification of variable nodes affects the structure of the graph as well as the value of check nodes, which in turn trigger the verification of other variable nodes in future iterations. The case in which the assigned value to a variable node is different from its true value, is called \textit{false verification}. If at least one of the distributions $f$ or $g$ is continuous, then false verification happens with probability zero \cite{SBB206,ZP08,ZP09}. Henceforth, we assume that the probability of false verification for a variable node in an iteration is zero. Using the Boole's inequality successively, it is easy to show that the probability of false verification in the whole recovery process is also zero.
%%%%%%%%%%%%%%%%%%%%%%%%%%%%%%%%%%%%%%%%%%%%%%%%%%%%%%%%%%%%%%%%%%%%%%%%%%%%%%%%%%%%%%%%%%%%%%%%%%%%%%%%%%%%%
%%%%%%%%%%%%%%%%%%%%%%%%%%%%%%%%%%%%%%%%%%%%%%%%%%%%%%%%%%%%%%%%%%%%%%%%%%%%%%%%%%%%%%%%%%%%%%%%%%%%%%%%%%%%%
\section{Message-Passing (MP) Interpretation of VB Recovery Algorithms}
\label{Decoding}
%%%%%%%%%%%%%%%%%%%%%%%%%%%%%%%%%%%%%%%%%%%%%%%%%%%%%%%%%%%%%%%%%%%%%%%%%%%%%%%%%%%%%%%%%%%%%%%%%%%%%%%%%%%%%
\subsection{Definitions and Setup}
A message-passing algorithm works in iterations through exchanging messages (belonging to certain alphabets) between check nodes and variable nodes along the edges in the graph, and processing the received messages at nodes by applying specific mapping functions. For the ease of presentation, we discuss the message alphabets and mapping functions only for the case in which the non-zero weights of the sensing graph are drawn from an uncountable or countably infinite alphabet set. For other sensing graphs, these parameters should be redesigned in order to have message-passing algorithms equivalent to the original algorithms discussed before.

Any message sent from a variable node belongs to an alphabet $\mathcal{M}:\{0,1\}\times \mathbb{R}$. The first coordinate, called \emph{status flag}, indicates the verification status of the variable node. The second coordinate represents the verified value of the variable node and is valid only if the status flag is $1$. Any message sent from a check node belongs to an alphabet $\mathcal{O}: \mathbb{Z}^{{}^{+}}\times \mathbb{R}$. The first coordinate indicates the number of unverified variable nodes neighbor to the check node. The second coordinate indicates the result of the linear combination of the unverified neighboring variable nodes. Moreover, the edges in NB-MP algorithms, multiply (divide) the second coordinate of a message by their associated weight, if it is sent from (received by) a variable node. Since the weights are chosen independently, all messages are independent.

Any iteration $\ell\geq 1$, consists of two rounds. A round starts with check nodes processing the received messages, then proceeds with the transmission of messages from check nodes to variable nodes, continues by variable nodes processing the received messages, and ends with the transmission of messages from variable nodes to check nodes. The two rounds in each iteration follow the same procedure and only differ in the mapping functions associated to variable nodes.

For the VB algorithms under consideration, the mapping functions in the variable nodes and check nodes are not functions of the iteration number. Hence, let $\Phi_v^{(1)},\Phi_v^{(2)}: \ma{O}^{d_v} \rightarrow \ma{M}$, represent the functions used at any \emph{unverified} variable node to map the incoming messages to the outgoing message in the first and second round of any iteration, respectively. When a variable node becomes verified at an iteration, its outgoing message remains unchanged, irrespective of its incoming messages. In contrast to the variable nodes, the mapping function used in check nodes is identical for both first and second round of each iteration. Every check node $i, i \in \{1,2,\cdots,m\} \triangleq [m]$ has an associated received measurement $\bs{c}_i$, a random variable taking values in $\mathbb{R}$. So, we use the notation $\Phi_c: \mathbb{R} \times \ma{M}^{d_c} \rightarrow \ma{O}$, to denote the mapping function used in all check nodes at any iteration. %%%%%%%%%%%%%%%%%%%%%%%%%%%%%%%%%%%%%%%%%%%%%%%%%%%%%%%%%%%%%%%%%%%%%%%%%%%%%%%%%%%%%%%%%%%%%%%%%%%%%%%%%%%%%
\subsection{Recovery Algorithms Based on Message-Passing}
\label{RecAlg}
Here we define the mapping functions $\Phi_v^{(1)}$, $\Phi_v^{(2)}$ and $\Phi_c$. Function $\Phi_v^{(1)}$ embeds the aforementioned verification rules DOCN and ECN, while function $\Phi_v^{(2)}$ embeds the ZCN rule. Let the message $\bs{o} \in \ma{O}$ be an ordered pair of elements $(d,c)$, where $d \in \mathbb{Z}^{^{+}}, c \in \mathbb{R}$, and the message $\bs{m} \in \ma{M}$ be an ordered pair of elements $(s,v)$, where $s \in \{0,1\}, v \in \mathbb{R}$. We also assume an arbitrary numbering for edges adjacent to a node (either variable node or check node), and use the notation $\bs{o}_{i}, i \in [d_v]$ and $\bs{m}_{j}, j \in [d_c]$.

We define the mapping functions $\Phi_v^{(1)}$, $\Phi_v^{(2)}$ embedding all three verification rules, discussed before.
\[
\Phi^{(1)}_v(\bs{o}_1, \cdots, \bs{o}_{d_v}) = \left\{
\begin{array}{ll}
	(1,c_i) & \exists i \in[d_v] : \bs{o}_{i} = (1,c_i)\\
	(1,c) & \exists i,j \in [d_v], i\neq j: \\
    & \bs{o}_{i} = (d_i,c) , \bs{o}_{j} = (d_j,c)\\
	(0,0) & \text{Otherwise}
\end{array}
\right.
\]

\[
\Phi^{(2)}_v(\bs{o}_1, \cdots, \bs{o}_{d_v}) = \left\{
\begin{array}{ll}
	(1,0) & \exists i \in [d_v] : \bs{o}_{i} = (d_i,0)\\
	(0,0) & \text{Otherwise}
\end{array}
\right.
\]

A variable node may not be verified to different values according to verification rules above for the same reasons
that the probability of false verification is zero.

For any iteration $\ell \geq 1$, the mapping function at any check node is as follows:
\[
\Phi_c(\bs{c}_i, \bs{m}_1, \cdots, \bs{m}_{d_c}) = (d_c - \ds_{i=1}^{d_c}{s_i},\bs{c}_i - \ds_{i=1}^{d_c}{s_i v_i}),
\]
where, $\bs{c}_i$ is the measurement associated with the check node $c_i$, and $\bs{m}_i = (s_i,v_i)$ is the message received along the $i$th edge. Since at iteration zero there is no incoming messages from variable nodes to check nodes, we have:
\[
\Phi^{(0)}_c(\bs{c}_i) = (d_c,\bs{c}_i).
\]
%%%%%%%%%%%%%%%%%%%%%%%%%%%%%%%%%%%%%%%%%%%%%%%%%%%%%%%%%%%%%%%%%%%%%%%%%%%%%%%%%%%%%%%%%%%%%%%%%%%%%%%%%%%%%
%%%%%%%%%%%%%%%%%%%%%%%%%%%%%%%%%%%%%%%%%%%%%%%%%%%%%%%%%%%%%%%%%%%%%%%%%%%%%%%%%%%%%%%%%%%%%%%%%%%%%%%%%%%%%
\section{Asymptotic Analysis Framework}
\label{analysis}
%%%%%%%%%%%%%%%%%%%%%%%%%%%%%%%%%%%%%%%%%%%%%%%%%%%%%%%%%%%%%%%%%%%%%%%%%%%%%%%%%%%%%%%%%%%%%%%%%%%%%%%%%%%%%
We first show that for given probability distribution functions $f,g$, and given constant parameters $d_v,d_c$ and $\alpha$ (i) the performance of a VB algorithm over realizations of the sensing graph and the input signal concentrates around the average performance of the algorithm, when the average is taken over all the elements in the ensemble $\ma{W}^{n}_f(d_v,d_c)\times \ma{V}^{n}_g(\alpha)$, and (ii) for $n \rightarrow \infty$, the average performance converges to the performance of the VB algorithm over a cycle-free graph defined as follows.\footnote{Our general method of proof is similar to that of \cite{RU01}. However, some non-trivial modifications are made since the message-passing interpretation of NB-VB algorithms lacks the extrinsic nature of standard message-passing algorithms of \cite{RU01}.} Let $\ma{N}_v^{2\ell}$, referred to as the neighborhood of node $v$ of depth $2\ell$, be the subgraph consisting of the variable node $v$ and all nodes (either variable node or check node) that are connected to $v$ with any path of length less than $2\ell$. A graph is cycle-free up to depth $2 \ell$ when for every $v$, $\ma{N}_v^{2\ell}$ is a tree.
%%%%%%%%%%%%%%%%%%%%%%%%%%%%%%%%%%%%%%%%%%%%%%%%%%%%%%%%%%%%%%%%%%%%%%%%%%%%%%%%%%%%%%%%%%%%%%%%%%%%%%%%%%%%%
\subsection{Concentration and Convergence to Cycle-Free Case}
Let $\beta^{(\ell)}$ be the fraction of variable to check node messages with unverified status in a general sensing graph at iteration $\ell$. Further, let $\mathbf{E}[\beta^{(\ell)}]$ denote the expected value of $\beta^{(\ell)}$, where the expectation is taken over all sensing graphs and input signals. Let $\alpha^{(\ell)}$ be the expected number of messages with unverified status passed along an edge emanating from a variable node with a tree-like neighborhood of depth at least $2\ell$ at the $\ell$th iteration, where the expectation is taken over all weights and all input signals. The following theorem shows that over all realizations, $\beta^{(\ell)}$ does not deviate far from $\mathbf{E}[\beta^{(\ell)}]$, which itself, is not far from $\alpha^{(\ell)}$, as $n \rightarrow \infty$, with high probability.

\begin{theorem}\label{Concentration}Over the probability space of all weighted graphs $\mathcal{W}^{n}_f(d_v,d_c)$, and all signal inputs $\mathcal{V}^{n}_g(\alpha)$, for fixed $\ell$, letting $\beta^{(\ell)}$ and $\alpha^{(\ell)}$ be defined as above, there exist positive constants $\mu(d_v,d_c,\ell)$ and $\eta(d_v,d_c,\ell)$, such that (i) for any $\epsilon>0$,
\begin{equation}
\label{EqDot}
\Pr\left[\left|\beta^{(\ell)}-\mathbf{E}[\beta^{(\ell)}]\right|>{\epsilon}/{2}\right]\leq 2e^{-\epsilon^2 n/\mu},
\end{equation}
and (ii) for any $\epsilon>0$, and $n>2{\eta}/{\epsilon}$,
\begin{equation}
\label{EqStar}
\left|\mathbf{E}[\beta^{(\ell)}]-\alpha^{(\ell)}\right|<{\epsilon}/{2}.
\end{equation}
\end{theorem}

Note that combining~\eqref{EqDot} and~\eqref{EqStar}, the following holds: for any $\epsilon>0$, and $n>2\eta/\epsilon$,
\[
\Pr\left[\left|\beta^{(\ell)}-\alpha^{(\ell)}\right|>{\epsilon}/{2}\right]\leq 2e^{-\epsilon^2 n/\mu}.
\]
Next, we show that $\alpha^{(\ell)}$ can be calculated under the assumption of cycle-free neighborhood for the SBB recovery algorithm using a deterministic procedure. A similar procedure exists for the LM algorithm, which is not discussed here due to the space limitation.
%%%%%%%%%%%%%%%%%%%%%%%%%%%%%%%%%%%%%%%%%%%%%%%%%%%%%%%%%%%%%%%%%%%%%%%%%%%%%%%%%%%%%%%%%%%%%%%%%%%%%%%%%%%%%
\subsection{State Parameters for the Analysis of the SBB Algorithm}
\label{notation}
In the analysis of NB-VB algorithms, the state parameters, at the beginning of any iteration $\ell$, are defined as the probability of variable nodes and check nodes belonging to some sets denoted by $\ma{K}^{(\ell)}$, $\ma{K}_i^{(\ell)}$, $\ma{R}^{(\ell)}$, $\Delta^{(\ell)}$, and $\ma{N}^{(\ell)}_{i,j}$. For instance, $\alpha^{(\ell)}$, described before as one of the state parameters, is indeed the probability that a variable node belongs to the set $\ma{K}^{(\ell)}$. In the following, we define the aforementioned sets for the SBB algorithm. For the LM algorithm, the definitions of the sets $\ma{K}_i^{(\ell)}$ and $\ma{N}^{(\ell)}_{i,j}$ change slightly. The set $\ma{K}^{(\ell)}$ consists of all unverified non-zero variable nodes, while the set $\Delta^{(\ell)}$ consists of all unverified zero-valued variable nodes. The set $\ma{R}^{(\ell)}$ includes all variable nodes recovered up to iteration $\ell$. Hence, $\ma{K}^{(\ell)}$, $\ma{R}^{(\ell)}$, and $\Delta^{(\ell)}$ are disjoint sets spanning the set of all variable nodes. We divide the set of all check nodes into subsets $\ma{N}^{(\ell)}_{i,j}$, where $i$ and $j$ indicate the number of neighboring variable nodes in the set $\ma{K}^{(\ell)}$ and $\Delta^{(\ell)}$, respectively. Further, the set $\ma{K}^{(\ell)}_{i}$ includes all variable nodes in $\ma{K}^{(\ell)}$ with $i$ neighboring check nodes in the set $\ma{N}^{(\ell)}_1:= \bigcup_{j=0}^{d_c-1} \ma{N}^{(\ell)}_{1,j}$.

In Theorem~\ref{SBBModel} below, we introduce the necessary and sufficient conditions for the verification of variable nodes in $\ma{K}^{(\ell)}$ in each iteration $\ell$ for the SBB algorithm. A similar theorem can be proved for the set $\Delta^{(\ell)}$ in each iteration $\ell$ for the SBB algorithm. Theorems with the same spirit as Theorem~\ref{SBBModel} can be proved also for the LM algorithm.

\begin{theorem}
\label{SBBModel}
In the first round of any iteration $\ell$ in the SBB algorithm, a non-zero variable node $v\in \ma{K}^{(\ell)}$ is verified if and only if it belongs to the set $\bigcup_{i=2}^{d_v}\ma{K}^{(\ell)}_i \cup \hat{\ma{K}}^{(\ell)}_1$, where the set $\hat{\ma{K}}^{(\ell)}_1$ consists of all variable nodes in the set $\ma{K}^{(\ell)}_1$ connected to the set $\ma{N}^{(\ell)}_{1,0}$.
\end{theorem}

Based on Theorem~\ref{SBBModel}, we can calculate the probability that a variable node in the set $\ma{K}^{(\ell)}$ is verified at iteration $\ell$, $\alpha^{(\ell)} - \alpha^{(\ell+1)}$, as a function of state parameters at the same iteration. Indeed, it can be shown that all state parameters at iteration $\ell$ are functions of the same parameters at iteration $\ell-1$. In the analysis we track the evolution of the state parameters with iteration. For an initial parameter $\alpha^{(0)}(=\alpha)$, the recovery algorithm is called successful if $\lim_{\ell \rightarrow \infty} \alpha^{(\ell)} = 0$, and is called unsuccessful if there exists $\epsilon > 0$, such that $\lim_{\ell \rightarrow \infty} \alpha^{(\ell)} > \epsilon$. The success threshold is then defined as the supremum of all $\alpha^{(0)}$, such that the algorithm is successful.
%%%%%%%%%%%%%%%%%%%%%%%%%%%%%%%%%%%%%%%%%%%%%%%%%%%%%%%%%%%%%%%%%%%%%%%%%%%%%%%%%%%%%%%%%%%%%%%%%%%%%%%%%%%%%
%%%%%%%%%%%%%%%%%%%%%%%%%%%%%%%%%%%%%%%%%%%%%%%%%%%%%%%%%%%%%%%%%%%%%%%%%%%%%%%%%%%%%%%%%%%%%%%%%%%%%%%%%%%%%
\section{Simulation Results}
\label{simulation}
Here, we present simulation results obtained by running the NB-VB algorithms over graphs of finite length $n$. We also present results obtained by running the proposed asymptotic analysis. The comparison of the results shows that there is a good agreement between empirical and analytical results for moderately large graphs ($n \approx 10^5$).

In all simulations, a signal element (variable node) belongs to the support set with probability $\alpha^{(0)}$. Also, each such element has a standard Gaussian distribution. The biregular graphs are constructed randomly with no parallel edges and all weights equal to one. In each experiment, the sensing graph is fixed and each simulation point is generated by averaging over 1000 random instances of the input signal.

For the analytical results, based on the fact that $\alpha^{(\ell)}$ is a non-increasing function of iteration number $\ell$, we consider the following stopping conditions:
\begin{enumerate}
	\item{Success:} $\alpha^{(\ell)} \leq 10^{-7}$.
	\item{Failure:} $\alpha^{(\ell)} > 10^{-7}$ and $|\alpha^{(\ell)} - \alpha^{(\ell-1)}| < 10^{-8}$.
\end{enumerate}
To calculate the success threshold, a binary search is performed until the separation between start and end of the search region is less than $10^{-5}$.

In Table \ref{success_threshold_1}, we have listed the analytical success thresholds of the LM and SBB algorithms for graphs with compression ration $r_c = d_c/d_v = 2$ and different $d_v$ values. For every graph, the SBB algorithm has better performance than LM. Also, as we decrease $d_v$, algorithms perform better in terms of success threshold and oversampling ratio $r_o = d_v/\alpha d_c$.\footnote{These results are consistent with the results observed for the Belief Propagation (BP) decoding algorithm of binary LDPC codes based on biregular graphs.} In fact, among the results presented in Table \ref{success_threshold_1}, the application of the SBB and LM to $(3,6)$ graphs results in the lowest oversampling ratio of $\approx 1.94$ and $\approx 2.94$, respectively.

\begin{table}[!h]
	\caption{Success Thresholds for different graphs and algorithms for fixed compression ratio $r_c = 2$}
	\begin{center}	
	\begin{tabular}{|l|c|c|c|c|c|}
		\hline
		$(d_v,d_c)$ & $(3,6)$ & $(4,8)$ & $(5,10)$ & $(6,12)$ & $(7,14)$\\
		\hline
		\hline
		 SBB& 0.2574 & 0.2394 & 0.2179 & 0.1992 & 0.1835\\
		\hline
		 LM & 0.1702 & 0.1555 & 0.1391 & 0.1253 & 0.1140\\
		\hline		
	\end{tabular}
	\end{center}
	\label{success_threshold_1}
\end{table}

\begin{figure}[!h]
\centering
\includegraphics[width=3.5 in]{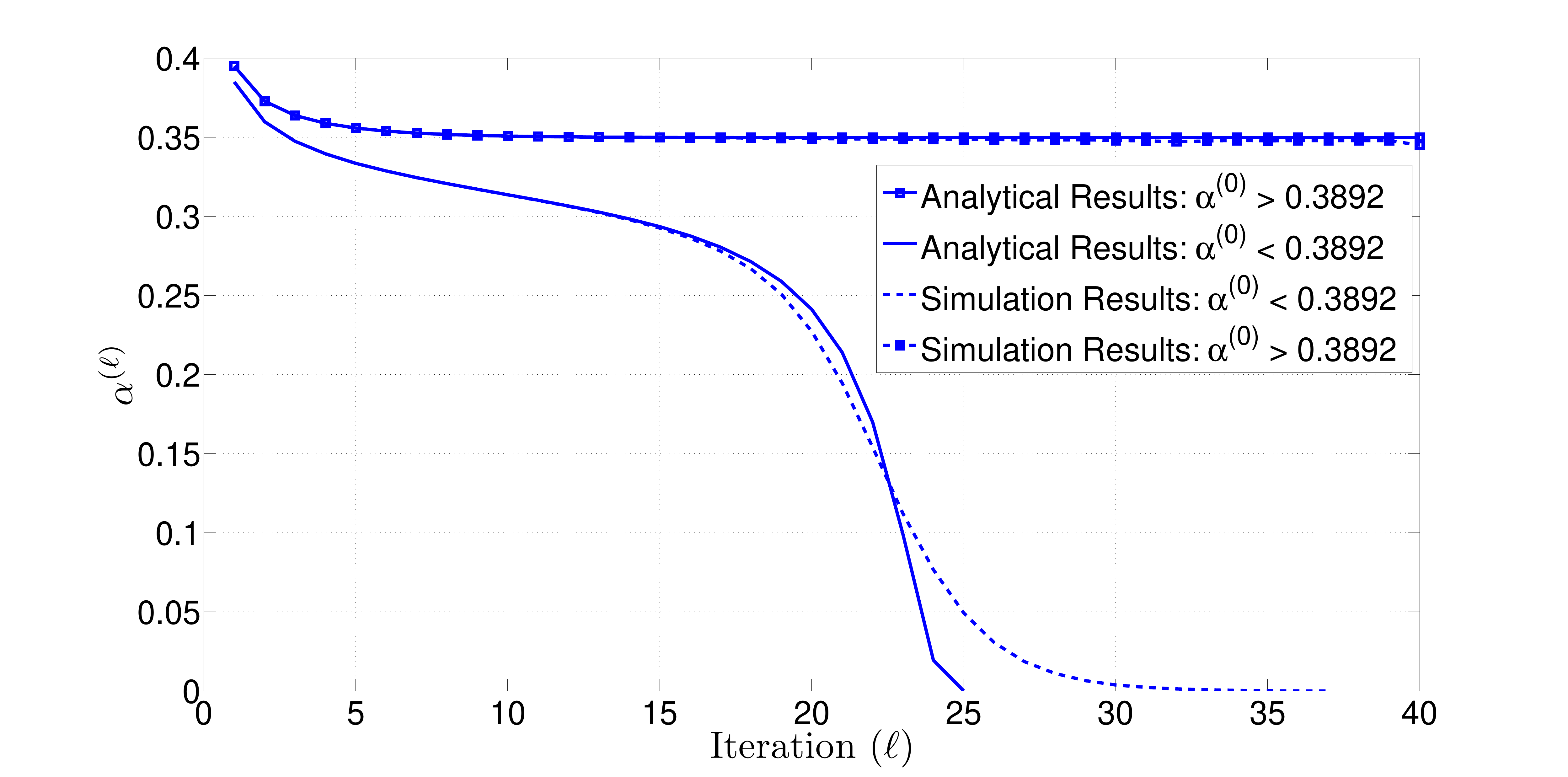}
\caption{Evolution of $\alpha^{(\ell)}$ vs. $\ell$ for the SBB algorithm over a $(5,6)$ graph.}
\label{G4G6Evolution100k}
\end{figure}

To further investigate the degree of agreement between our asymptotic analysis and finite-length simulation results, we have presented in Fig.~\ref{G4G6Evolution100k} the evolution of $\alpha^{(\ell)}$ with iterations $\ell$ for the SBB algorithm over a $(5,6)$ graph with $n=10^5$. Two values of $\alpha^{(0)}$ are selected: one above and one below the success threshold, which is $0.3892$. The theoretical results are shown by solid lines while simulations are presented with dashed lines. The two sets of results are in close agreement particularly for the cases where $\alpha^{(0)}$ is above the threshold and for smaller values of $\ell$.

\begin{figure}[!ht]
\centering
\includegraphics[width=3.5 in]{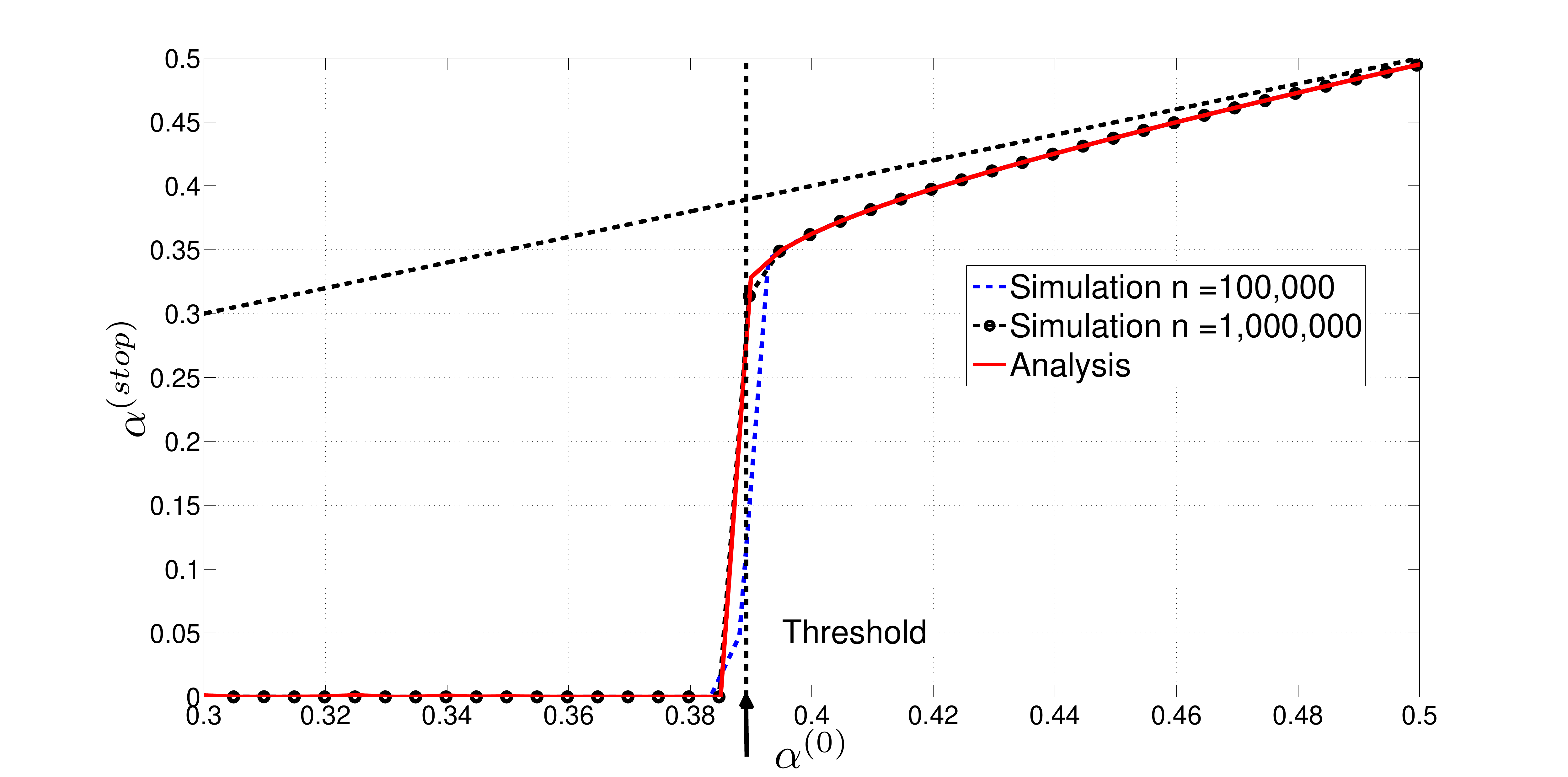}
\caption{$\alpha^{(stop)}$ vs. $\alpha^{(0)}$ for SBB applied to $(5,6)$ graphs.}
\label{RecoveredVarNodes}
\end{figure}

Next, for different values of $\alpha^{(0)}$, we estimate the average fraction of unverified non-zero variable nodes $\alpha^{(\ell)}$ using the analysis, and denote the value of $\alpha^{(\ell)}$ at the time that the analysis stops (because one of the stopping criteria is met) as $\alpha^{(stop)}$. These values are plotted vs. the corresponding values of $\alpha^{(0)}$ in Fig. \ref{RecoveredVarNodes} for the SBB algorithm over the $(5,6)$ sensing graphs. In the same figure, we have also given the corresponding simulation results for two randomly selected $(5,6)$ sensing graphs with $n = 10^{5}$ and $10^{6}$. The simulation results for both lengths closely match the analytical results, with those of $n = 10^{6}$ being practically identical to the analytical results. From the figure, it can also be seen that as $\alpha^{(0)}$ increases and tends to one, the curves tend to the asymptote  $\alpha^{(stop)} = \alpha^{(0)}$.

%\setcounter{figure}{0}
%\vspace{5cm}

As the last experiment, we compare the running time and accuracy of the proposed asymptotic analysis against those in the differential equation approach presented in \cite{ZP07J}. For comparison, a biregular $(3,6)$ graph and the SBB algorithm were chosen. 
%The binary search for the success threshold starts with interval $[0.2,0.3]$ and ends when the separation between the start and end region in less than $10^{-5}$. 
The analysis is implemented in MATLAB and executed on an AMD Phenom 9650 Quad-Core 2.3 GHz CPU 
with 3 GB RAM. The success threshold of $0.2574$ is obtained in $23.1$ seconds. 
Table \ref{pcompare} summarizes the results of running the analysis of~\cite{ZP07J} on the same machine
for different values of $n$. The reported thresholds increase with the increase in $n$. 
For $n=100,000$, the running time is roughly 100 times that of our proposed method.
Moreover, the obtained threshold of $0.2591$ is only in agreement with the threshold of $0.2574$,
obtained by the proposed method, up to two decimal points. In fact, experiments similar to those reported in
Fig.~\ref{G4G6Evolution100k} reveal that the accuracy of the threshold obtained by the method of~\cite{ZP07J}
is lower than our results. In particular, our simulations show that the SBB algorithm over 
$(3,6)$ graphs with $n=10^5$ fails for $\alpha^{(0)} = 0.259$, which would imply that the threshold $0.2591$
is only accurate up to two decimal points. 
\begin{table}[!h]
	\caption{Success threshold and running time of the differential equation analysis of \cite{ZP07J}.}
	\begin{center}	
	\begin{tabular}{|l|c|c|}
		\hline
		$n$ & Success Threshold & Running Time (seconds)\\
		\hline
		$1,000$ & 0.2577 & 9.9 \\
		\hline
		$10,000$ & 0.2589 & 103.9 \\
		\hline
		$20,000$ & 0.2590 & 220.6 \\
		\hline
		$50,000$ & 0.2590 & 647.4 \\
		\hline
		$100,000$ & 0.2591 & 2044.1 \\
		\hline
	\end{tabular}
	\end{center}
	\label{pcompare}
\end{table}
%%%%%%%%%%%%%%%%%%%%%%%%%%%%%%%%%%%%%%%%%%%%%%%%%%%%%%%%%%%%%%%%%%%%%%%%%%%%%%%%%%%%%%%%%%%%%%%%%%%%%%%%%%%%%
%%%%%%%%%%%%%%%%%%%%%%%%%%%%%%%%%%%%%%%%%%%%%%%%%%%%%%%%%%%%%%%%%%%%%%%%%%%%%%%%%%%%%%%%%%%%%%%%%%%%%%%%%%%%%
%\vspace{-10pt}
\bibliographystyle{IEEEtran}
\bibliography{Journal_biblio}
%%%%%%%%%%%%%%%%%%%%%%%%%%%%%%%%%%%%%%%%%%%%%%%%%%%%%%%%%%%%%%%%%%%%%%%%%%%%%%%%%%%%%%%%%%%%%%%%%%%%%%%%%%%%%
%%%%%%%%%%%%%%%%%%%%%%%%%%%%%%%%%%%%%%%%%%%%%%%%%%%%%%%%%%%%%%%%%%%%%%%%%%%%%%%%%%%%%%%%%%%%%%%%%%%%%%%%%%%%%
\end{document}